# Angle resolved Photoemission from Ag and Au single crystals: Final state lifetimes in the *as* range

F. Roth[1,2], T. Arion[2], H. Kaser[3], A. Gottwald[3], and W. Eberhardt[2,4]

[1] TU Bergakademie Freiberg, Institute of Experimental Physics, Leipziger Straße 23, D-09599 Freiberg, Germany
[2] Center for Free Electron Laser Science (CFEL), DESY, Notkestr. 85, D-22607 Hamburg, Germany
[3] Physikalisch-Technische Bundesanstalt (PTB), Abbestr. 2-12, D-10587 Berlin, Germany
[4] Inst. for Optics and Atomic Physics, TU Berlin, Strasse des 17. Juni 135, D-10623 Berlin, Germany

We present angle resolved photoemission spectra for Ag(111) and Au(111) single crystals in normal emission geometry, taken at closely spaced intervals for photon energies between 8 eV and 160 eV. The most dominant transitions observed are attributed to f-derived final states located about 16 eV to 17 eV above $E_F$ for both materials. These transitions exhibit very distinct resonance phenomena and selection rules and are reminiscent of the angular momentum characteristics of the states involved. The excited electron lifetime is in the attosecond (*as*) range as determined from the energy width of the observed transitions. This serves as an alternate approach to the direct determination of excited electron lifetimes by *as* laser spectroscopy.

**Introduction**
Angle resolved photoemission using a continuously tunable excitation source, such as monochromatized synchrotron radiation, has been developed over the past decades [1][2][3][4][5][6][7][8][9] and become the experimental method of choice, when the momentum resolved energy band structure in solids, including their surfaces, is to be experimentally determined throughout the entire Brillouin zone with high precision. While for periodic solids two momentum components and the energy of the states are unambiguously defined via energy and momentum conservation rules, the third momentum component, oriented perpendicular to the surface needs to be derived at by other means. The most simplistic zero order approximation is based on the assumption of free electron like final states, usually modified by an inner potential to account for the momentum change in passing through the surface, and/or a modified electron mass. Outside the crystal in vacuum and in the immediate surface region these are the same states that are observed when scattering a free electron beam by the periodic surface and that is why they are also referred to as inverse LEED final states [10][11][12][13][14]. Inside the crystal these states however have to be matched to the periodic unoccupied Bloch states existing throughout the crystal.

As next step of refinement, the transitions occurring at the critical points of the band structure can be identified, usually by searching for extreme values in the dispersion of the initial state bands. Once such an extreme value has been observed, the final state of the transition is localized using the photon energy at which the spectrum is measured. This has been demonstrated and systematically explored for the first time in determining the critical points of the band structure of Ni [15]. Once these points of the final state band structure have been identified, again a free electron like band can be fitted to interpolate between the exactly determined critical points. Through monitoring the intensity of the direct interband transitions it is also possible to gain information about the imaginary part of the self-energy of the final states of the transitions, which is commonly referred to as the lifetime of the excited electron in the final state of the transition [16]. This has been recently extensively documented for Cu and it was found that distinct interband transitions can be identified up to photon energies well above 100 eV photon energy. While the energy of the finals states in Cu can be reasonably well approximated by free electron like final states, the lifetime of the electrons in these bands does not match the predictions of a free electron like final state [17].

Here we extend these studies to Ag and Au single crystals each with a (111) surface orientation. For both these metals quite intense transitions have been observed quite early in differential reflectance spectroscopy [18] as well as in the early experimental and theoretical photoemission work [19][20][21][22][23]. These studies revealed transitions, observed for photon energies between 15 eV and 25 eV, which were attributed to final state bands with atomic 4f or 5f character, and these are definitely not free electron like bands. Here we have studied the angle resolved photoemission spectra in the normal direction with much more detail and over a large photon energy range, in order to address the very large intensity variations in the coupling to these f-derived bands. Monitoring the width of the direct transitions observed over the entire photon energy range enables us to get an approximation of the lifetime of the excited electrons in the final conduction band states as a function of energy. These lifetimes are in the attosecond (*as*) range and similar to some recent results obtained by *as*-laser spectroscopy using high harmonic generation (HHG) source [24][25][26].

**Experimental**

The photoelectron spectra were recorded with the SCIENTA R4000 hemispherical electron spectrometer in the iDEEAA end station [27] at the Metrology Light Source (MLS) of the Physikalisch-Technische Bundesanstalt in Berlin, Germany [28]. The highly polarized (about 99.5 % p-polarisation) radiation was delivered by the Insertion Device Beamline (IDB) with an U125 undulator (125 mm period length) as the source. The monochromator combines normal incidence (NI) with grazing incidence (GI) diffraction geometry and provided monochromatized radiation from approximately 1.5 eV to 10 eV in the NI mode and 10 eV to 280 eV in the GI mode. The typical photon flux is about some $10^{12}$ photons/s and the resolving power is better than 900 (GI mode) and 2500 (NI mode). The spot size on the sample is about 1.7 mm horizontal and 0.1 mm vertical. The spectra were taken with the electron spectrometer at a pass energy of $E_{pass} = 20$ eV, resulting in an electron energy resolution of 25 meV. The angular resolution was chosen to be ± 0.14 degrees by integrating the signal over the appropriate number of channels on the detector.

**Sample preparation and orientation**

The samples were prepared by several cycles of argon-ion bombardement (ion energy $E = 500$ eV) and annealing to approximately 550°C for several minutes. Afterwards, the cleanliness and order of the surface was verified by measuring the dispersion of the well known surface states of the (111) surfaces. The sample preparation took place in the preparation chamber (base pressure smaller than $2 \times 10^{-6}$ Pa) direct connected to the analysis chamber of the iDEEAA apparatus (base pressure about $5 \times 10^{-8}$ Pa). The SCIENTA electron spectrometer is mounted such that it takes a full angular spectrum of ± 7 degrees in the horizontal direction, whereas in the perpendicular direction we had to rely upon the mechanical mounting of the crystal and the manipulator [27]. Measuring the surface state dispersion curves of the (111) oriented crystal surfaces we estimate the vertical angle to be aligned with an accuracy better than 0.5 degrees. The angle of incidence of the light was 45° and the polarization was p-polarized in the plane of detection of the analyzer. This polarization enables us to observe all direct transitions possible in normal emission geometry as postulated by the selection rules [29][30].

## Results --- Normal emission intensity variations as function of photon energy

The normal emission geometry plays a special role in angle resolved photoemission, since for this geometry the final state of the photoelectron and the photoemission matrix element both have to be completely symmetric with respect to all symmetry operations of this crystal axis [29][30]. This puts a very convenient constraint on the selection of possible final states contributing to the photoemission process in this geometry inside the crystal. In the surface region these states map onto the free electron states propagating into the vacuum, the so-called inverse LEED states [10][11][12].

Fig. 1 displays the normal emission photoemission intensity distributions as a function of binding energy and excitation (photon) energy in a linear false color representation for Ag(111) and Au(111). The well established surface state is visible as a faint feature at low binding energies near $E_F$, whereas the most prominent emission features arise from the emission of d-symmetry bands at binding energies between 4 eV and 7 eV for Ag and 2 eV and 6.5 eV for Au. The larger bandwidth of the d-band in Au is due to the increase in spin-orbit splitting of the bands, otherwise the observed features are quite similar. In this kind of presentation the similarities in the general evolution of the photoemission with change in photon energy are quite stunning. What is also immediately apparent are the very strong transitions observed around 20 eV to 25 eV in photon energy. These transitions are so strong that we had to readjust the scaling of the false color plots in Fig. 1. at an excitation energy of 30 eV. Otherwise the transitions at higher excitation energies are hardly visible at all. As judged by the binding energy of the observed bands, these transitions reflect the Γ-point of the band structure. This is also roughly in agreement with the assumption of free electron-like final states. These transitions had been observed before. The earliest reference is in early photoemission results [31] and differential reflectivity measurements [18]. Theory was also very well advanced concerning these transitions [32] and in general these are attributed to 4f- and 5f-symmetry related unoccupied final state bands.

The cross section for the d-bands falls off quite rapidly over this large photon energy range, which is an atomic feature. The 4d-states in Ag as well as the 5d-states in Au display a so called Cooper minimum [33] of the photoexcitation cross section around a photon energy of 100 eV (Ag) and 150 eV(Au) [34].

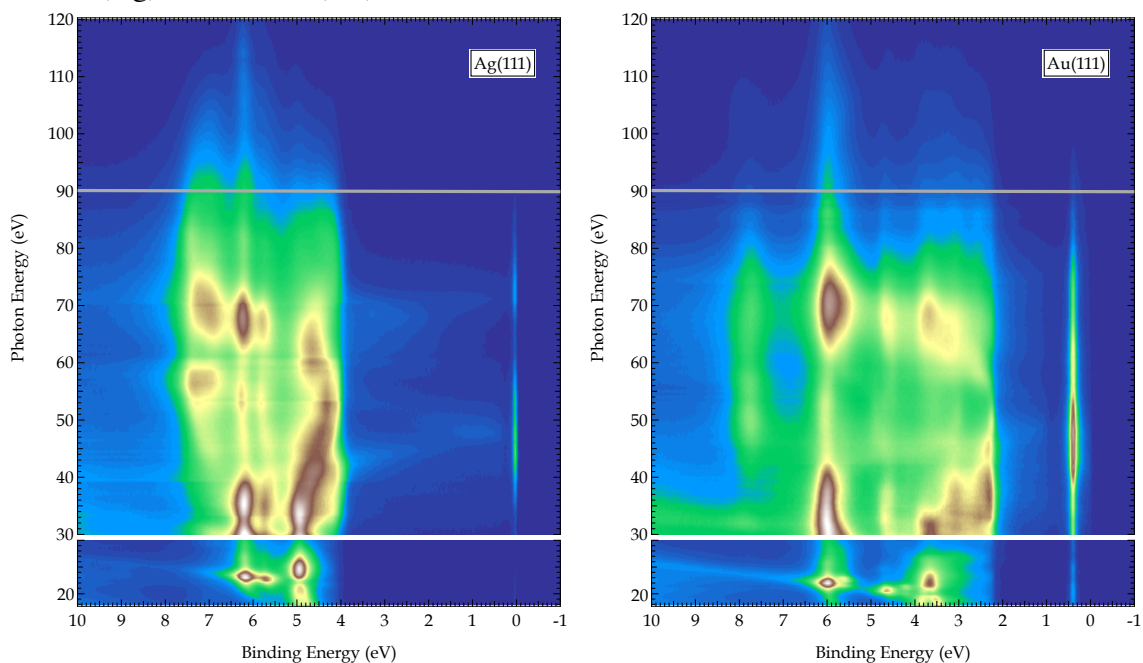

Fig. 1 Normal emission intensity distributions for Ag(111) (left) and Au(111) (right), taken for photon energies between 16 eV and 120 eV. (To enhance the visibility of all features throughout the whole photon energy range we readjusted the scaling of the false color plot at an excitation energy of 30 eV – marked with a white, horizontal line).

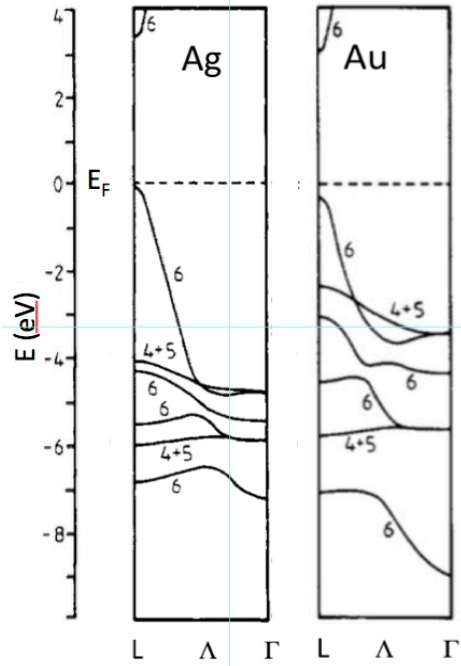

Fig. 2 Calculated relativistic valence bands of Ag (left) and Au (right) along the Λ-axis from Γ to L in the Brillouin zone (adapted from ref. [35]). These are the initial bands that are probed in normal emission in the (111) surface orientation.

The calculated (relativistic) band structure [35] along the Λ-axis from Γ to L in the Brillouin zone for Ag and Au is shown for comparison in Fig. 2. There are earlier band structure calculations by Christensen and Janak et al. [36][37][38], which show the essentially the same features and band energies. We have chosen to show these calculations [35], since they also include the full symmetries of the bands and critical points (see also Fig. 5) which will be important for the discussions later.

In this surface orientation the Γ-Point is indeed reached at a final state energy of about 20 eV for Ag and 18 eV for Au, referenced relative to the Fermi level $E_F$. By visual inspection of Fig. 1, the next Γ-point is reached at a final state energy of 62 eV for Ag and 65 eV for Au. This is far from the estimates of a free electron like final state approximation, even taking into account an effective mass increased by about 12 % for Ag and 14 % for Au as suggested earlier [39].

The maximum in the emission intensity of the surface state emission has been assigned to reflect the L-point of the final state band structure [39]. This is observed approximately for a photon energy of 46 eV for both Ag and Au, whereas the Ag emission intensity exhibits a pronounced shoulder about 8 eV higher photon energy than the main peak what is also in agreement with previous data [39][40]. The corresponding intensity profiles for both the Ag(111) and Au(111) surface state emissions are shown in Fig. 3. What has not been reported so far is the subsequent maximum in the emission intensity of the surface state at a photon energy of about 72 eV. This higher energy maximum is especially prominent for Ag, while it is also observed as a weaker structure for Au. At the same photon energies the initial bulk bands are observed at energies, which suggests to locate the corresponding transitions near the Γ-point of the bulk band structure (see above). While it is always permissible to speculate about surface Umklapp processes or secondary Mahan cones, the prominent maximum of the surface state emission at these final state energies clearly awaits an in depth theoretical calculation since it contradicts all previously discussed empirical and theoretical models for the intensity variations of the surface state emission [39][41].

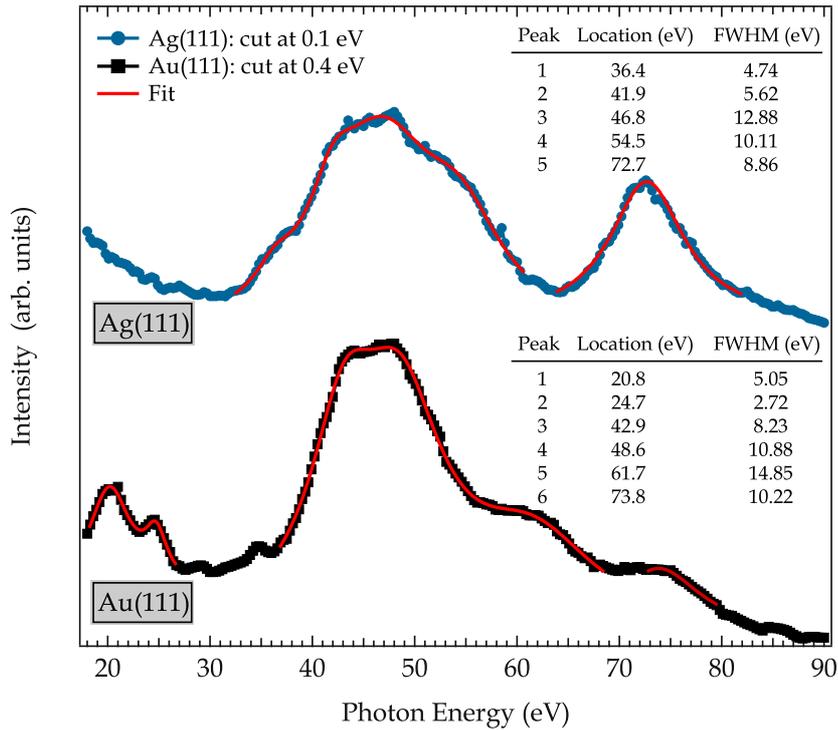

Fig. 3 Intensity of the surface state photoemission as a function of photon energy for both the Ag(111) and the Au(111) surface. These curves can be partially approximated by a series of Lorentzians, the peak positions and widths (FWHM) are listed in the tables. The red curves show the result of these locally fitted sections.

Let us now turn the discussion to another very prominent feature in the spectra shown above, the extremely strong transitions observed for photon energies between 18 eV and 30 eV (cf. Fig. 1). This energy range is shown expanded in Fig. 4 and we also included in this figure the corresponding spectra for Cu, which already have been published earlier [17]. Again the similarities are quite stunning. The d band emissions display an extraordinarily strong resonance behavior over a very narrow photon energy range, while the s-p symmetry surface state does not show these strong resonances. Moreover, the emission of the topmost d-band displays an anti-resonance at the exact final state energy where the lower d-band emission goes through a maximum. This final state energy varies for the three materials and is indicated by the dashed white lines in Fig. 4.

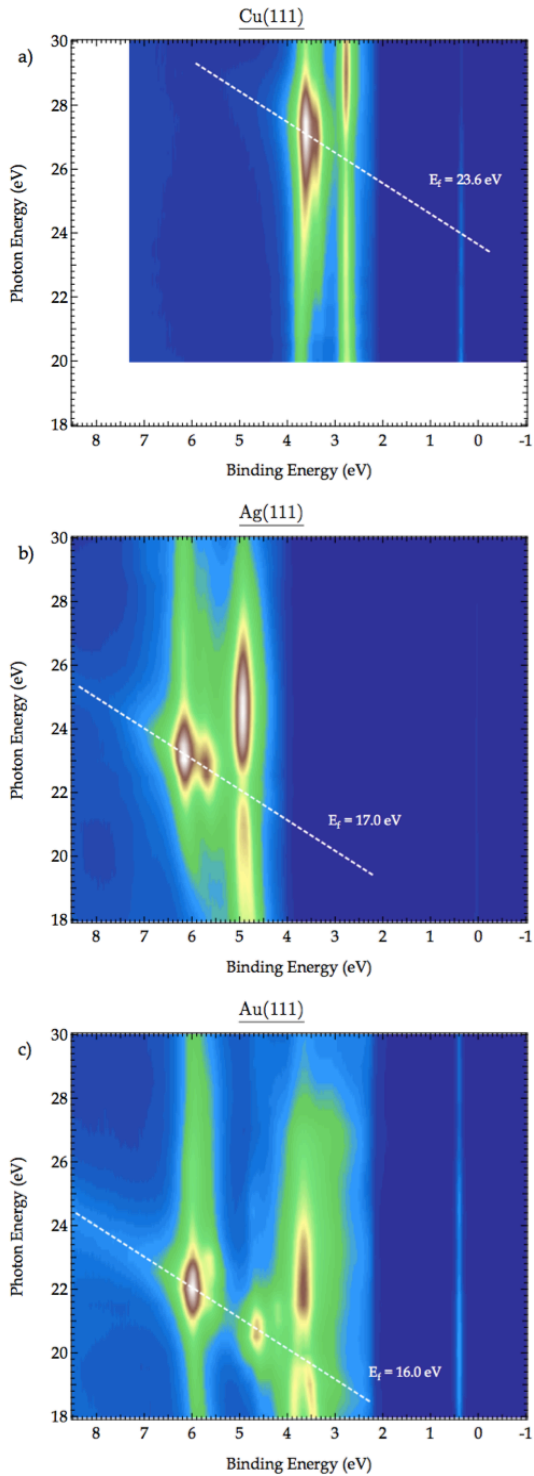

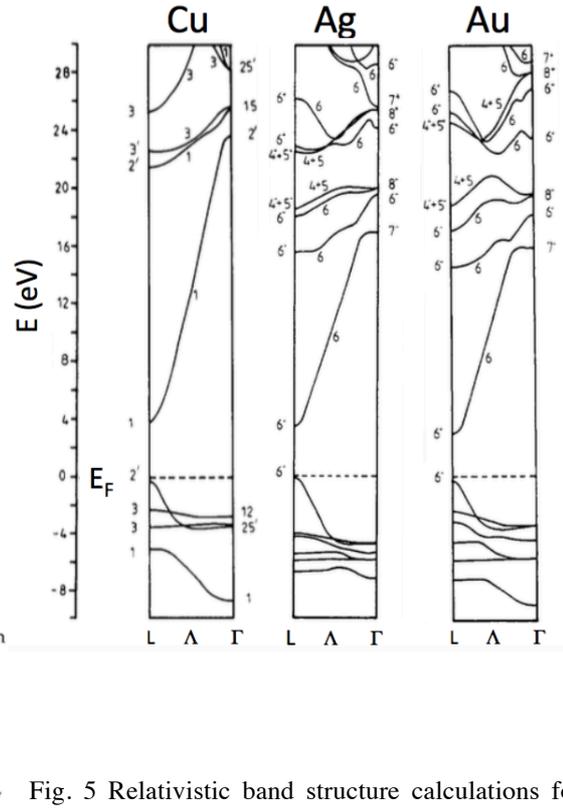

Fig. 5 Relativistic band structure calculations for Cu, Ag, and Au along the Λ symmetry line of the bulk band structure showing not only the initial states, but also the final states and symmetries up to energies 30 eV above $E_F$. The bands and critical points for Ag and Au are labeled using the relativistic notations, while the symmetry labels for Cu are non-relativistic (adapted from [35]).

Fig. 4 False color photoemission intensity maps of the normal emission from Cu(111), Ag(111), and Au(111) displaying the extremely strong resonances in the d-band emissison for photon energies between 18 eV and 30 eV. The white dashed lines mark the final state energy, where the higher binding energy d-band emission is strongest. The observed initial state energies correspond to the band energies at the Γ-point of the bulk band structure.

Starting from low photon energies, there is only one final state band of $\Lambda_1$-($\Lambda_6$) symmetry, that enables normal emission photoemission [29][30], whereas the $\Lambda_3$-symmetry band (and the corresponding relativistic bands) does not contribute to normal emission. Since we are taking spectra under an angle of incidence of 45 degrees in p-polarisation, both $\Lambda_1$ and $\Lambda_3$ initial state bands are allowed for transitions into this final band. As the photon energy is increased the direct transitions approach the Γ-point of the band structure. At the Γ-point, however, the symmetry is higher than along the Λ-axis of the Brillouin zone. At Γ itself transitions between $Γ_{2,5'}$ and $Γ_{2'}$ are allowed, whereas transitions between $Γ_{1,2}$ and $Γ_{2'}$ are forbidden by symmetry. For $Γ_{1,5}$ final state symmetry both, transitions from $Γ_{2,5'}$ and $Γ_{1,2}$ initial states are allowed. This increased symmetry at Γ thus is the reason for the observed anti-resonance of the lower binding energy d-band emission displayed prominently for all three elements in Fig. 4. In the relativistic case these symmetry arguments also hold, since a forbidden transition remains essentially forbidden. Even though the selection rules might be somewhat relaxed, the transition matrix elements remain zero or negligibly small unless the relativistic interaction substantially changes the character of the bands [32]. For an allowed transition, on the other hand, generally the corresponding transition intensities (matrix elements) are redistributed between the bands split by the spin-orbit interaction.

For Cu this symmetry argument and the resonance vs. anti-resonance in emission intensity was not discussed before, whereas for Ag and Au the resonance and anti-resonance of the d-band emission has been reported quite extensively in previous studies. It was first noted by Wehner et al. [31] and later again by many authors [32][42][43][44][45]. The first detailed discussion of these drastic changes in the emission intensity were presented by Benbow and Smith [32] based upon numerical approximations of the transition matrix elements. Even though these were not first principles calculations, the matrix elements show that the transition near Γ from the lower $Λ_3$-band to the lowest energy $Λ_1$-band is more than an order of magnitude stronger than the transition to the second unoccupied band. For the next higher occupied $Λ_3$-band the matrix element at Γ to the lower final state band vanishes, whereas it has a value of about 20 % of the strong matrix element from the lower $Λ_3$-band for transitions to the second unoccupied band. This essentially reflects our observations presented in Fig. 4 to the extent that the lower part of the d-band emission couples strong with the lowest energy unoccupied final state band, whereas the upper part of the d-band couples preferentially to the second unoccupied final state band. The basic symmetry argument presented above that due to an increase in symmetry at Γ the transitions from the upper d-band to the lowest final state band are not allowed, were lost in these numerical exercises.

While the general parameterized matrix element calculations reflect the behavior of the experiment, Wern et al. [42] used these matrix elements, calculated by Benbow and Smith [32] and additionally invoked surface transmission factors for the two relevant final state bands. These decreased the emission involving the first unoccupied $Λ_1$-band near Γ, while abruptly raising the emission from the next higher $Λ_1$-band in the immediate vicinity of Γ. This additional transmission factor however does not improve the understanding or explain the resonance vs. anti-resonance of the transitions. It is clearly demonstrated by the data in Fig. 4, that exactly the same final state gives rise to this phenomenon. All subsequent studies [43][44][45] simply referenced this publication [42] without adding additional insight into this particular point.

The existence of quite dominant final states in the band structure of Ag and Au about 16 eV to 17 eV above $E_F$ was also noted in optical reflectivity studies [18] as well as electron scattering experiments [46]. In this context it is interesting to note that the surface state emission is not significantly modulated by these final state bands. These are bulk bands and, at least for Ag and Au, they exhibit very noticeable modifications due to relativistic effects. For Ag and Au this has been attributed to the admixture of 4f -and respectively 5f- atomic character unoccupied states. In this context the atomic angular momentum selection rules seem to survive in the solid, even though the angular momentum is not a valid quantum number for delocalized electronic states in a solid. The transitions to these final states are predominant for d-like states, whereas the s-p-like bands and the s-p-like surface state show negligible coupling.

**The width of the final state bands and excited electron lifetimes**

As in our previous study of the final states in Cu [17], we aim to extract the energy widths of the transitions and relate these to the lifetimes of the excited electrons in these states. This is complementary information to the direct measurements of electron lifetimes by laser excited two-photon photoemission. Using HHG sources, these studies also are capable to address a similar range of final states [24][47]. As long as we restrict our analysis to transitions near the critical points of the band structure, both the initial and final state bands are fairly flat and thus any momentum uncertainty due to the finite angular acceptance of our spectrometer becomes negligible. For other transitions, both the slope (group velocity) of the initial state as well as of the final states have to be corrected for as described in [16]. Assuming a free electron-like final state for this correction represents an upper limit, since the slope of the real final state bands in general is smaller than the slope of a free electron-like final state band. The group velocity of a band is given by

$$\delta E/\delta \mathbf{k} = 2E/\mathbf{k}$$

Accordingly the broadening due to the finite angular resolution can be expressed as

$$\Delta E = 2E\, \Delta \mathbf{k}\, /\mathbf{k} = 2E \sin(\delta\Theta).$$

For an angular resolution of $\delta\Theta = \pm 0.14°$ this relates to $\Delta E = 0.01E$.

This means that the maximum linewidth contribution due to the finite angular resolution and group velocity of the final state bands is about 1 % of the final state total energy. This slope, or group velocity, scales approximately as the momentum vector in the extended zone scheme ---- i.e., the slope in the 3rd Brillouin zone is three times as large as in the first Brillouin zone.

Most of these curves display a strong maximum near 20 eV photon energy and a second one around 65 eV to 70 eV. As discussed above these transitions correspond to initial states near the Γ-point of the bulk band structure. For Ag(111) in Fig. 6 we only plotted curves corresponding to the direct transitions around the Γ-point of the band structure. For Au(111) in Fig. 7 we also include transitions from the topmost d-band near the L-point in Fig. 7a. This assignment of the location of the transition in momentum space immediately follows from the selection of the initial state energy (2.3 eV). For these states the emission is strongest near a photon energy of about 35 eV. Technically the curves shown in Figs. 6 and 7 are derived from vertical cuts at selected initial state energies in Fig. 1. The initial state energy or the energy of these cuts is listed in each panel of Figs. 6 and 7. For a direct interband transition between a unique initial and final state, we expect these intensity profiles shown in Figs. 6 and 7 to be described quite well by a single Lorentzian centered at the appropriate transition energy. This is repeated for each energy where these transitions are occurring. Superimposed on this is a general drop in intensity upon increasing the photon energy as dictated by the evolution of the overall photoelectron cross section [34].
The location of these Lorentzians and their widths (FWHM) are shown in the form of inserted tables in each of the panels of Figs. 6 and 7. The red lines in these figures show the results of these (local) fits of the transition intensities.

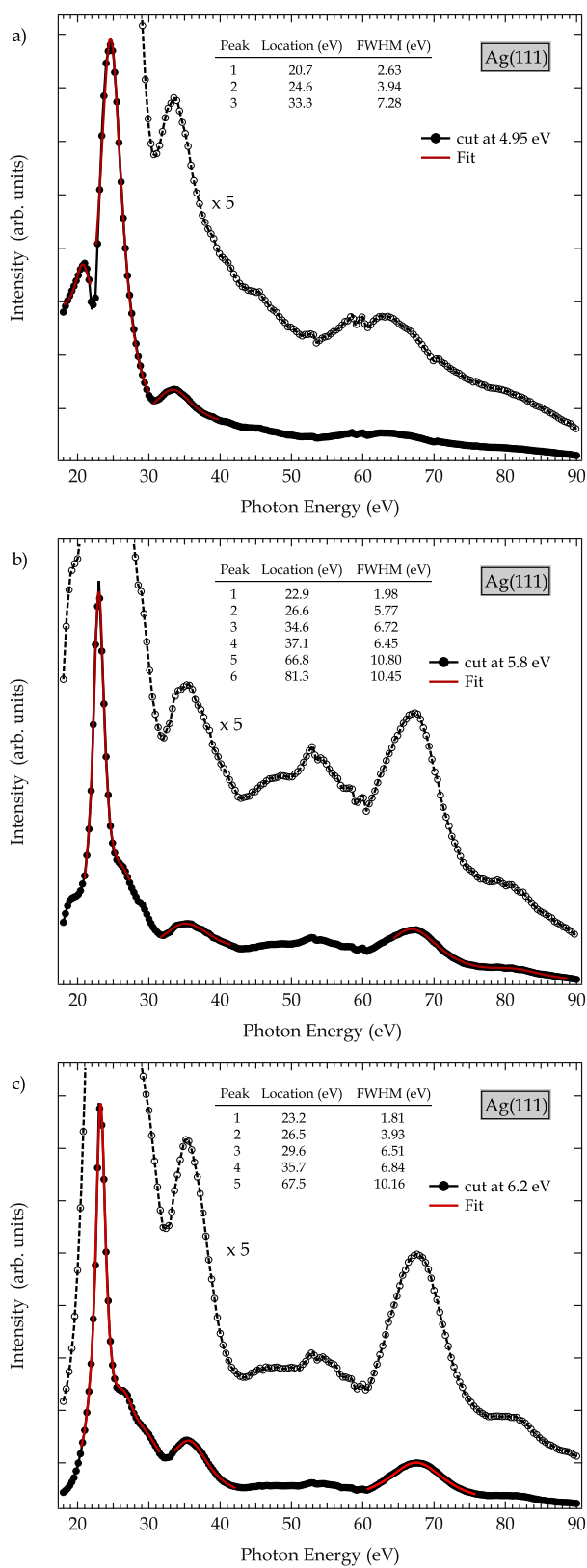

Fig. 6 Intensity profiles of normal emission d-band features of Ag(111) at selected binding energies (panels a-c) derived from cuts along vertical lines in Fig. 1. These intensity profiles are fitted locally by Lorentzians as indicated by the red sections of the curves. The positions and widths of these Lorentzians are also included in tabular form in each of the panels.

Resonant photoemission contributions arising from the decay of a 4p core hole have a threshold value of 57.6 eV for the Ag $4p_{3/2}$ level and 57.2 eV for the $5p_{3/2}$ core hole in Au (NIST database)[*]. Via a super Coster-Kronig Auger decay these core holes could cause an enhancement of the photoemission lines in a so-called resonant photoemission process. The peaks detected in our spectra with maxima around photon energies of 67 eV to 68 eV, however are far above these values and therefore not significantly caused by this resonant process. Similarly the Auger decay of a 4f-hole in Au – the $4f_{7/2}$ binding energy is 84 eV – also does not lead to significant enhancements of the direct interband transitions observed in the intensity profiles of Fig. 7.

---

[*] https://srdata.nist.gov/xps

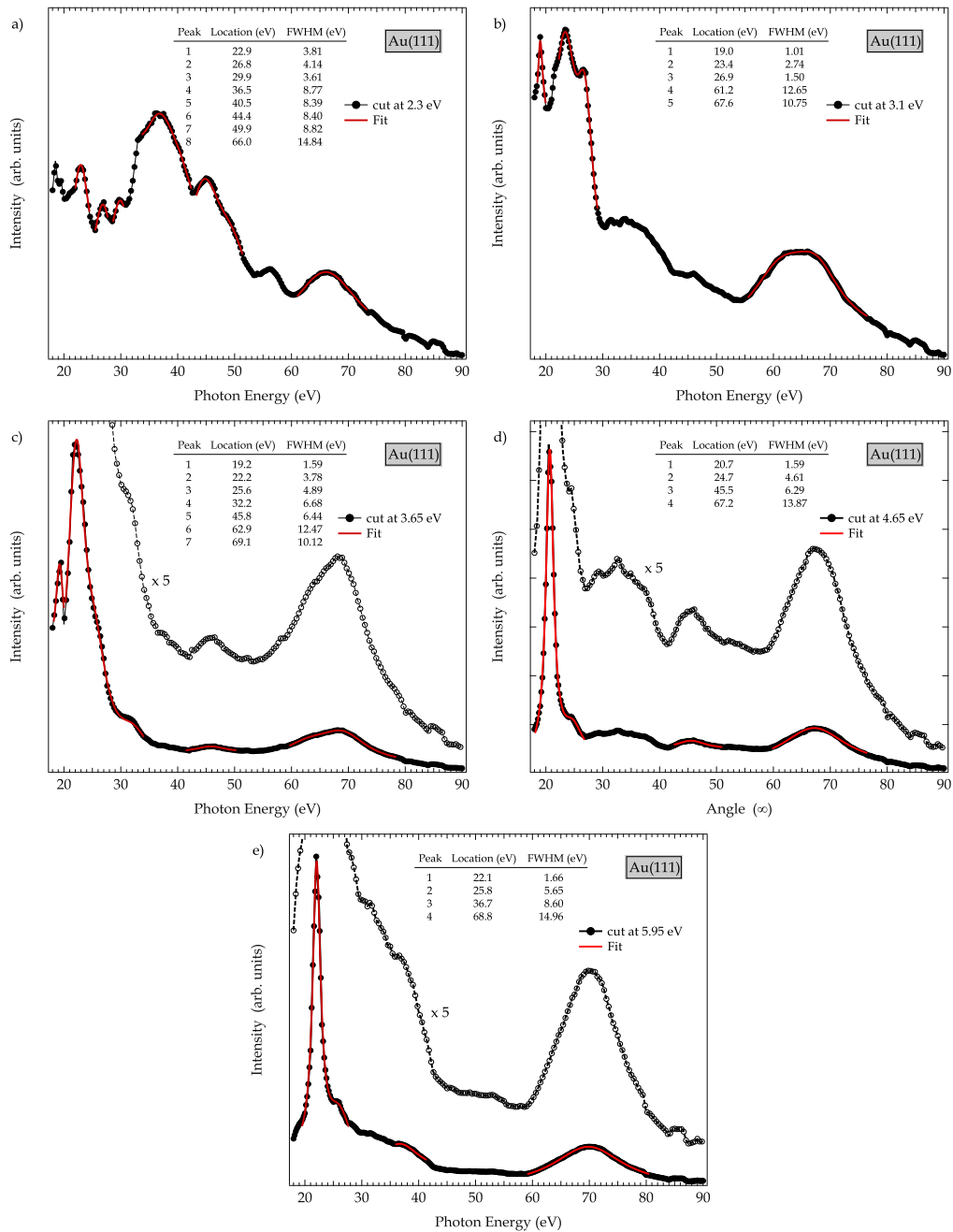

Fig. 7 Intensity profiles of normal emission d-band features of Au(111) at selected binding energies (panels a-e) derived from cuts along vertical lines in Fig. 1. These intensity profiles are fitted locally by Lorentzians as indicated by the red sections of the curves. The positions and widths of these Lorentzians are also included in tabular form in each of the panels.

Additional contributions to these widths arise from the initial state lifetime, i.e., the Auger decay lifetime (width) of the hole in the valence band. Recently there were some calculations published investigating these lifetimes at least partially [47][48]. As soon as the hole is within the d-band the calculated hole lifetimes decrease substantially. The calculated lifetime broadening due to electron-electron scattering for a hole at an energy of -5 eV (relative to $E_F$) is about 50 meV for Ag and about 100 meV for Au [48]. In addition there is a substantial contribution to the lifetime due to electron-phonon scattering which is surprisingly calculated to be larger, i.e., 100 meV for Ag and 150 meV for Au. Brown et al. [47] present similar results. They calculate total lifetimes, including both electron-phonon and electron-electron scattering, for a hole at these energies of 1.5 fs for Au and 2 fs for Ag. This puts the broadening to about 220 meV for Au and 165 meV for Ag.

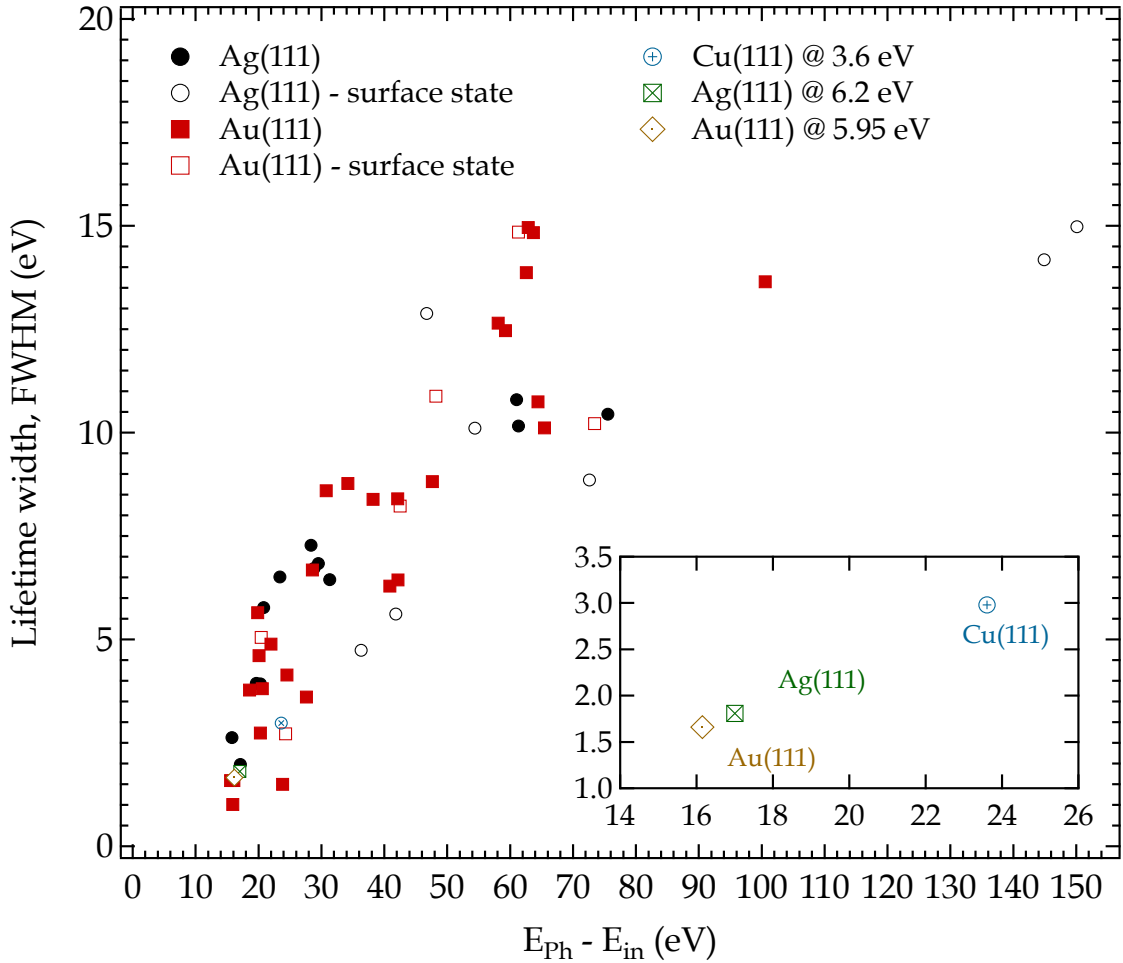

Fig. 8 Summary of the measured widths of the Lorentzians approximating the direct interband transitions for the surface states (Fig. 3) as well as the bulk band structure transitions (Figs. 6 and 7) observed in normal emission from the Ag(111) and Au(111) surfaces. The inset panel shows the width of the overall most dominant transitions involving the final states near the Γ-point of the second Brillouin zone.

Electron-phonon scattering of the excited electron in the final state also contributes to the widths of the observed transitions. Although there are no calculations available, we can reliably conclude from previous temperature dependent measurements on Cu [49] that for our measurements, which were taken at room temperature, electron-phonon scattering of the final state excited electron does not contribute significantly to the broadening of the Lorentzians in the fits of Figs. 6 and 7.

In summary, we conclude that on the scale of the measured transition energy widths, with the exception of the lowest transitions below 20 eV, the widths of the Lorentzians (as determined from the curves shown in Figs. 3, 6, and 7) reflect the final state lifetime broadening of the excited electrons in Ag and Au. All these widths are compiled in graphical form in Fig. 8. The hole lifetime, including possible phonon broadening effects, is also directly visible in the electron distribution curve (EDC) of the valence band photoemission. We obtain EDC's via horizontal cuts in our data shown in Figs. 1 and 4. We observe in these data directly that the energy width of the structures in the EDCs (horizontal widths) are approximately a factor of 5 to 10 smaller than the width of the same features along the transition energy scale (vertical axis). This is immediate proof for our assumption that the hole lifetime and electron-phonon scattering both do not significantly enhance the width of the observed transitions and the data which are summarized in Fig. 8.

The electron lifetime widths shown in Fig. 8 are steeply increasing from about 1.5 eV to 8 eV between 15 eV and 40 eV final state energy. At higher energies these values are still increasing even though with a much smaller slope. Scattering calculations for a free electron gas also predict this steep rise [50], but the curve exhibits a broad maximum between 50 eV and 60 eV and then a slight decrease in values for 80 eV and higher. This also reflects the electron mean free path, which is smallest around 50 eV to 60 eV in energy and then starts to gradually increase. This is straightforward to understand, since the electron mean free path is given by the product of the electron lifetime in the final state and its group velocity. Similar to our results for Cu published earlier [17], the widths measured here do not reflect a decrease in electron scattering rates, but rather a slow increase even for energies of 80 eV and above. We have to admit here however that we have only very few data points for direct transitions in Ag and Au, whereas the corresponding data were much more abundant for Cu.

Concentrating on the strongest transitions observed, shown as inset in Fig.8, we observe a width of 1.6 eV for Au, 1.9 eV for Ag and 3.0 eV for Cu. Making use of the energy-time uncertainty relation

$$(\Delta E)*(\Delta t) \geq \hbar/2,$$

we can get an estimate of the excited electron lifetimes at these energies. The numerical lifetime values thus derived are 205 *as* for Au, 175 *as* for Ag, and 110 *as* for Cu. These estimated values have to be taken with great caution. We do not even want to enter into the discussion whether the energy-time uncertainty relation is strictly valid. There seems to be sufficient evidence that this is a good estimate. For example the widths of core electron lines observed in XPS and the core hole decay times derived from these widths, which have lead to the core-hole clock method [51], make good sense. However there are even more objections/restrictions to discuss. The interpretation of the energy width of the transition and the assumption of a Lorentzian lineshape depends on the concept of a transition between stationary states of the system. The smaller the times are the more it has to be questioned whether this is a valid approximation. Additionally, even the absorption of the photon is not instantaneous but takes a finite time as pointed out by Wigner [52].

Recently the lifetime of the excited electrons in Ni, Ag, and Au at about 25 eV excitation energy has been determined in two photon photoemission experiments using *as*-HHG laser sources [24][26]. An overview of technique using *as*-streaking in photoemission from solids has been summarized by Cavalieri et al. [25], while the original developments were carried out using gas phase spectroscopy of atoms. The *as*-streaking or RABBIT technique is based upon superposition of an IR laser pulse with the HHG-*as*-pulse on the target and observing the electrons that actually have absorbed one photon each from the two laser fields as a function of the phase shift between the two laser fields. Tao et al. [24] thus derived a value for the lifetime in the order of >200 *as* for an excited electron at 25 eV in Ni, whereas Locher et al. [26] also took the transition time for the photon absorption into account in their interpretation and modeling of the data. They derived a combined lifetime/transition time of these states in Ag and Au with values between 200 *as* and 100 *as*, even though some of the derived values were negative. This is due to the complex modeling that tries to take the Wigner time into account as well. However, we find it reassuring that without this complex modeling, their raw data show clear sidebands at a delay of 100 *as* between the HHG pulse and the IR pulse, whereas at a delay of 800 *as* these sidebands are missing, i.e., the excited electron has vanished from the conduction band states populated in the initial photon absorption process.

**Conclusions**

We presented angle resolved photoemission intensity maps taken in the direction of the surface normal of a Ag(111)- and a Au(111)-oriented crystal. For both these single crystalline metals the initial state electronic structure has been profoundly well characterized in the past, but the data presented here aim at shining new light at the final states participating in the photoemission process and the electron scattering rate (or lifetime, respectively) in these final states. The lifetime data are also complementary to very recent experiments, which aim at measuring the excited electron lifetimes in final states by *as* laser sources making use of HHG energy conversion.

As previously observed for Cu, for a final state energy up to at least 100 eV, a multitude of clear and well-defined final states are observed in these photoemission spectra. For Ag and Au, however, the by far most prominent final states are observed near the Γ-point of the second Brillouin zone at energies of 16 eV for Au and 17 eV for Ag, which are attributed to final state bands with 4f- (Ag) and 5f- (Au) atomic character. This atomic character is corroborated by the fact that these final state bands show large relativistic splittings, which also increase substantially from Ag to Au. Moreover, the s-like surface state exhibits negligible coupling to these final states. The strong variations in transition intensity for various initial states coupling to these final states can be explained using the dipole selection rules for these transitions. The major factor here is that the increased symmetry at the Γ-point as compared to the Λ-axis of the Brillouin zone results in a forbidden transition right at the Γ-point between some of the bands, whereas the transition is allowed all along the Λ -axis. This results in very dramatic intensity changes for the emission of the topmost d-states. This interpretation is supported by the fact that the exactly analogous behavior is also observed in angle resolved Cu(111)-photoemission at a final state energy of about 24 eV.

We observed well-defined direct transitions between initially occupied band states and unoccupied final states inside the crystal and at its surface. Inside the crystal, these states are given by the band structure of the bulk periodic crystal, and in the surface region there exist and contribute additional surface states and resonances as well as evanescent free electron states from the vacuum. This entire photoemission process can now be described by the proper advanced theoretical calculations and we are looking forward to a full theoretical analysis of our data by state of the art calculations.

**Acknowledgment**

We thank the staff of the Metrology Light Source of the PTB, for experimental and technical support. This research was funded through the research program 'Structure of matter' of the Helmholtz Association of research centers in Germany.